\documentclass{article}

\usepackage{arxiv}

\usepackage[utf8]{inputenc} 
\usepackage[T1]{fontenc}    
\usepackage{hyperref}       
\usepackage{url}            
\usepackage{booktabs}       
\usepackage{amsfonts}       
\usepackage{nicefrac}       
\usepackage{microtype}      
\usepackage{lipsum}		
\usepackage{graphicx}
\usepackage{natbib}
\usepackage{doi}
\usepackage{tabularx}
\usepackage{amsmath}
\usepackage{multirow}
\usepackage{float}
\usepackage{caption}
\usepackage{graphicx}
\usepackage{array}
\usepackage{rotating}
\usepackage{textcomp}
\usepackage{xcolor}
\usepackage{subfig,epsfig,tikz}
\usepackage{booktabs}
\usepackage{siunitx}
\usepackage[toc,page]{appendix}
\usepackage{titlesec}

\title{Dynamic pattern of imbalance states between
resting-state functional brain networks -during
development}

\author{ \hspace{1mm}Fahimeh Ahmadi  \\
	Physics Department\\
	Shahid Beheshti University \\
	1983969411, Tehran - Iran \\
	\texttt{} \\
    \And
    \hspace{1mm}Zahra Moradimanesh  \\
	The Institute for Cognitive and Brain Sciences\\
	Shahid Beheshti University \\
	1983969411, Tehran - Iran \\
	\texttt{} \\
    \And
    \hspace{1mm}Reza Khosrowabadi  \\
	The Institute for Cognitive and Brain Sciences\\
	Shahid Beheshti University \\
	1983969411, Tehran - Iran \\
	\texttt{} \\
	\And
	\hspace{1mm}G.Reza Jafari \\
    Physics Department, Shahid Beheshti University \\ 1983969411, Tehran - Iran \\
	\texttt{gjafari@gmail.com} \\
}

\renewcommand{\headeright}{}
\renewcommand{\shorttitle}{Patterns of imbalance states between sub-brain regimes during development in the resting state}

\hypersetup{
pdftitle={A template for the arxiv style},
pdfsubject={q-bio.NC, q-bio.QM},
pdfauthor={David S.~Hippocampus, Elias D.~Striatum},
pdfkeywords={First keyword, Second keyword, More},
}

\begin{document}
\maketitle

\begin{abstract}
The functional brain network emerges from the complex, coordinated activity of distinct yet connected regions, which underlie the diverse repertoire of human cognitive functions. Structural Balance Theory (SBT) has been successfully applied to model such nontrivial connections through the analysis of balance and unbalance triadic configurations. In this study, using SBT, we examine the network of imbalanced triads in the resting-state brain subnetworks, which undergo dynamic changes during development. We demonstrate that anticorrelation patterns evolve across the lifespan, reflecting a developmental trajectory from a locally modular organization in childhood to a flexible and reconfigurable architecture during adolescence and finally to a highly segregated and functionally specialized network system in adulthood. This developmental trajectory indicates that the spread of anticorrelations is not an inherent feature of brain organization. This mature organization facilitates a balance between self-referential, internally generated cognitive processes and externally oriented, goal-directed cognition, enabling efficient and adaptive cognitive control. This balance is underpinned by prominent anticorrelations between the Default Mode Network (DMN) and the Frontoparietal Network (FPN)  in adulthood. This is while during adolescence these anticorrelations are substantially weaker, suggesting that the maturation of these network connections from adolescence to adulthood establishes a functional architecture that supports the segregation of internal and external cognitive processes.  These findings elucidate how the dynamic evolution of anticorrelation patterns in brain networks supports cognitive development across the lifespan, offering new insights into the neural basis of adaptive cognitive control.
\end{abstract}

\keywords{Network Neuroscience\and Structural Balance Theory (SBT) \and resting-state functional MRI (rs-fMRI)}

\flushbottom
\maketitle
%
%
\thispagestyle{empty}

\section*{Introduction}
Both social and natural systems typically manifest as complex networks composed of interconnected components \cite{albert2002statistical}, and the brain is a remarkable example \cite{sporns2004organization}, comprising approximately 100 billion neurons \cite{fornito2016fundamentals}, each capable of forming thousands of synaptic connections with other neurons \cite{shepherd2003synaptic}. These intricate connections enable communication between neurons, allowing the brain to operate as a complex network characterized by two primary forms of connectivity: Structural connectivity refers to the brain's physical wiring via axons and white matter tracts \cite{vskoch2022human}, while functional connectivity reflects the dynamic interactions between regions based on correlated activity patterns. Unlike structural links, functional connectivity captures how regions communicate, often studied using functional MRI (fMRI) as a proxy for neural activity \cite{rogers2007assessing}.

The application of graph theory and network science to the study of functional brain connectivity was pioneered by researchers such as Olaf Sporns \cite{sporns2004motifs}, who introduced graph-theoretical methods to map and analyze the brain's functional connectivity. Other seminal contributions were made by Edward Bullmore \cite{hagmann2008mapping}, who highlighted the small-world properties of brain networks, and Michael G. Watts and Steven H. Strogatz \cite{watts1998collective}, whose small-world network model provided the foundation for understanding the efficiency and modularity of brain connectivity. Numerous studies have investigated the brain using network science approaches. This framework has been widely applied in both healthy and clinical populations, offering insights into typical development, aging, and neurological disorders \cite{dickerson2009large,cohen2020brain,haqiqifar2025patterns}.
Recent studies have applied structural balance theory (SBT) to brain networks, revealing that complex interactions beyond pairwise connections significantly shape their dynamics. 

Research on disorders such as autism spectrum disorder (ASD) has highlighted atypical triadic interactions, emphasizing the importance of SBT in understanding functional connectivity, especially in neurodevelopmental disorders \cite{moradimanesh2021altered}.
\\
SBT also explores the role of negative links, or anti-correlated interactions, which are rare but crucial for maintaining network stability. Their collective behavior, though non-trivial, has a profound impact on network balance and reveals the complexity of brain regions. This study demonstrated that negative functional connections in the resting-state brain network form hubs, creating a uniquely signed network topology that drives the system toward more balanced states with lower energy costs \cite{saberi2021topological}. Neuroplasticity refers to the brain’s ability to modify and adapt, a well-established concept defined as the neural system’s capacity for change \cite{schaefer2017malleable,kelly2014strengthening}. Many studies highlight its decline across the lifespan. Changing demand is a key factor in neural adaptation, and one study used SBT to examine the requirement for change of the functional brain network. They proposed that the number of triadic frustrations—equal to the number of imbalanced triads—indicates how much the brain network needs to change. Their findings revealed a U-shaped pattern of change requirement over the lifespan, with the highest demand in early and late life stages and the lowest in adults. They then focused on the brain’s negative subnetwork, due to the role of negative links in forming frustration, and found a strong correlation between negative link density and the number of frustrations. The negative subnetwork exhibited a distinct topology, following a log-normal degree distribution with a specific peak. Moreover, global features such as hubness, efficiency, and clustering changed noticeably in adulthood. Based on these findings, they inferred that the highly plastic brain in early life adapts to high demands and supports functional development. In adulthood, reduced demand coincides with lower but sufficient neuroplasticity. In contrast, the aging brain’s diminished plasticity cannot meet the increasing demands—likely rooted in structural degeneration—leading to impaired functionality and cognitive decline \cite{saberi2021requirement}.


It is known that a negative link between two brain regions in the functional network indicates the presence of anti-synchrony. Previous studies have shown that anti-synchronous interactions play an important role in the brain and influence the balance of its network. Individual negative links, or anti-synchronies, are important, but two fundamental questions arise:
First, what is the underlying pattern in the placement of these anti-synchronies during lifespan? In other words, we aim to examine whether anti-synchronies spread out from childhood to adulthood and exert global effects across the brain network, or whether they remain a localized phenomenon. To answer this, we must consider higher-order interactions, as they contain structural information that cannot be captured by pairwise connections alone. Structural Balance Theory (SBT) serves as a suitable framework for understanding such network structures. Second, what does this pattern look like at the scale of brain subnetworks? To address this, we rescale our analysis from individual brain regions to subnetworks, since brain subnetworks are known to contain meaningful cognitive and behavioral information. We demonstrate that anti-synchrony between two brain regions can indeed affect the broader brain network, especially the interactions between subnetworks. Moreover, the extent of this influence varies across different age groups.
\\

\section*{Materials and methods}
\subsection*{Participants}
This study utilized a dataset comprising resting-state functional magnetic resonance imaging (rs-fMRI) data from healthy individuals across different developmental stages. The current version of the dataset was released in March 2021. The raw imaging and demographic data were originally collected and shared through the publicly available Autism Brain Imaging Data Exchange I, accessible at\href{https://fcon_1000.projects.nitrc.org/indi/abide/abide_I.html}{ABIDE I}
\\A total of 215 participants were included in the final dataset, distributed across three age groups: childhood (7–11 years old, n = 58), adolescence (12–17 years old, n = 97), and adulthood (18–30 years old, n = 60). Data were selected exclusively from ABIDE I sites where participants were instructed to rest with their eyes open while fixating on a projected cross. This protocol was chosen due to its enhanced reliability in resting-state analyses, as it reduces the likelihood of participants falling asleep during scanning.
To minimize potential confounding effects of biological and behavioral variability, only right-handed male individuals were included. Further inclusion criteria required that each participant had at least five minutes of usable rs-fMRI data following preprocessing. All participants were fully verbal and had an IQ greater than 70, assessed using the Wechsler Abbreviated Scale of Intelligence (WASI and/or WASI-IV). Individuals with any prior or concurrent diagnosis of neurological disorders (e.g., epilepsy, meningitis, encephalitis, head trauma, or seizures) were excluded.
\paragraph{Ethical considerations.}The present study used anonymized, open-access data from the ABIDE I repository. All data were originally collected under institutional review board (IRB) approval at each contributing site, and informed consent was obtained from all participants. No new data were collected, and no identifying information was used in this study.

\subsection*{Data Preprocessing}
We used preprocessed resting-state functional magnetic resonance imaging (rs-fMRI) data from healthy individuals obtained from the Autism Brain Imaging Data Exchange I (ABIDE I) database, which is publicly available through the Preprocessed Connectomes Project (PCP) at \href{http://preprocessed-connectomes-project.org/abide/}{ ABIDE preprocessed}. The current version of the dataset was released in March 2021.
Among the four available preprocessing pipelines, we selected data processed using the Configurable Pipeline for the Analysis of Connectomes (CPAC). The following preprocessing steps were applied:
\begin{enumerate}
    \item \textbf{Basic preprocessing}: This included realignment, slice timing correction, and spatial registration (boundary-based rigid body (BBR) alignment for functional-to-anatomical registration, and ANTs for anatomical-to-standard (MNI152) registration). Signal intensity normalization was performed using a 4D global mean of 1000.
    
    \item \textbf{Nuisance regression}: To remove confounding signals due to head motion, physiological noise, and scanner drift, we regressed out 24 motion parameters (six motion parameters, their squares, and first-order derivatives), tissue signals using the CompCor method, and linear and quadratic trends.
    
    \item \textbf{Motion scrubbing}: Volumes with framewise displacement (FD) greater than 0.5 mm or global signal changes exceeding 3 standard deviations were scrubbed. Only participants with a minimum of 150 valid time points (approximately 5 minutes of data) were retained for further analysis.
    
    \item \textbf{Global signal}: The global mean signal was not regressed out during preprocessing.
    
    \item \textbf{Spatial smoothing and filtering}: Spatial smoothing was applied using a Gaussian kernel with a full width at half maximum (FWHM) of 6 mm. Band-pass filtering (0.027–0.073 Hz; Slow-4 band) was performed during the denoising step, as this frequency band has been associated with greater reliability in detecting resting-state networks.
    
    \item \textbf{Quality assessment}: The denoised data were visually inspected using quality assurance (QA) plots from the Conn toolbox.
    
    \item \textbf{Site harmonization}: To control for site-related variability across the five ABIDE I acquisition sites (NYU, SDSU, UM, USM, and Yale), the ComBat harmonization method was applied. Specifically:
    \begin{itemize}
        \item For each participant, the lower triangle of the $400 \times 400$ functional connectivity matrix was vectorized to form a $79{,}800 \times 1$ feature vector.
        \item These vectors were concatenated column-wise to produce a $79{,}800 \times 215$ data matrix, where each column corresponds to one participant.
        \item A batch vector ($215 \times 1$) was created to represent site codes (from 1 to 5).
        \item A model matrix ($215 \times 2$) containing age and group labels (all healthy controls) was used to preserve biological variation during harmonization.
        \item Kruskal–Wallis tests (FDR-corrected at a 5\% significance level) were performed before and after harmonization to assess site effects. Initially, 1.37\% of connections showed significant site effects, which dropped to zero after ComBat adjustment.
    \end{itemize}
\end{enumerate}

\subsection*{Construction of the brain networks from resting-state fMRI data}
In this study, we defined the nodes of whole-brain networks based on 400 regions of interest (ROIs) as introduced by the 2-mm Schaefer-400 atlas\cite{schaefer2018local}. For the analysis of brain subnetworks, we used the 17-network Yeo parcellation defined within the same atlas. In particular, within this parcellation, the default mode network (DMN) and the salience network (SN) are represented by 32 and 34 ROIs, respectively.
To define the edges of the whole-brain network, we computed functional connectivity as the Pearson correlation between all pairs of the 400 ROIs for each participant, using the entire resting-state blood-oxygen-level-dependent (BOLD) time series. This resulted in 215 individual weighted correlation matrices of size 400 × 400. The same approach was applied to construct functional connectivity matrices for each of the 17 Yeo functional subnetworks.
To stabilize the variance and facilitate statistical analysis, we applied Fisher’s z-transformation to the Pearson correlation coefficients (r), converting them into normally distributed z-values. For each edge (i, j), we conducted two-tailed one-sample t-tests on the Fisher z-transformed values across participants to assess whether the mean correlation coefficient differed significantly from zero. Edges that did not survive the 5\% significance level (after Bonferroni correction for multiple comparisons) were set to zero. No additional manual thresholding was applied to the networks to avoid introducing arbitrary biases into the network topology.
All computations were performed using MATLAB and the Statistics Toolbox (Release 2017b). Brain networks were visualized using BrainNet Viewer (version 1.63)\cite{xia2013brainnet}.

   \subsection*{Balance theory}
\begin{figure}[H]
    \centering
    \includegraphics[width=0.6\linewidth]{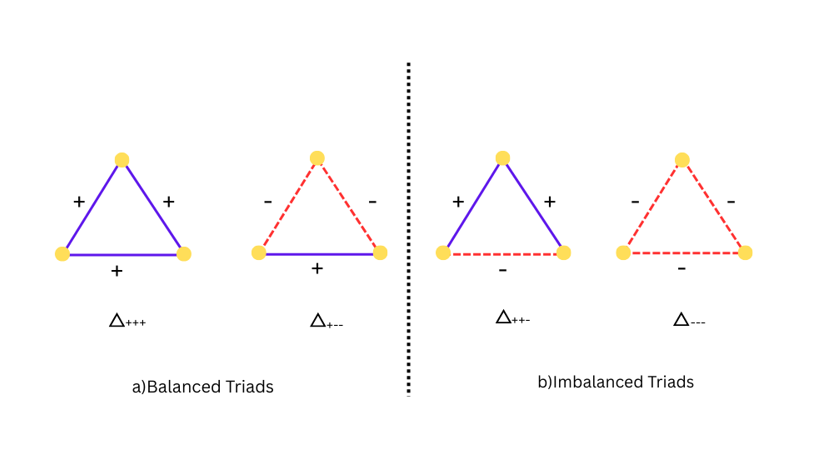}
    \caption{\textbf{Illustration of the four possible triad types defined by Structural Balance Theory (SBT): balanced ($\Delta_{+++}$), balanced ($\Delta_{+--}$), imbalanced $\Delta_{++-}$, and fully imbalanced $\Delta_{---}$} }
    \label{fig:enter-labelee}
\end{figure}

 The concept of balance theory was first introduced by Heider in 1946 in the field of social psychology. The theory considers triadic relationships; in other words, the theory states that when two individuals are in a friendship, the introduction of a third person who is in a relationship with both of them has an impact on the relationship between the two. Heider assigned positive (negative) sign value to pairwise interactions in the context of friendship (enmity) relationships. A balanced (imbalanced) relationship is defined as one in which the product of the signs is positive (negative). \cite{heider1946attitudes, heider2013psychology} In this context, four types of triads are defined: balanced ($\Delta_{+++}$),  balanced ($\Delta_{+--}$), imbalanced ($\Delta_{++-}$), and imbalanced ($\Delta_{---}$) Fig.\ref{fig:enter-labelee}.
 
 A fundamental tenet of the balance theory is that certain structures are in a balanced state, while others are not \cite{heider2013psychology,wasserman1994social}. It is important to note that an imbalanced triad represents a frustrated configuration, as it creates tension within the network and drives its links to adjust in a way that resolves the imbalance and restores a balanced state \cite{saberi2021topological}. In the real world, people are involved in many tripartite relationships at the same time, so balance theory should apply to the set.
This work was first done by Cartwright and Harari. They developed the notion into the field of graph theory, which was subsequently named the Structural Balance Theory (SBT) \cite{cartwright1956structural}. Heider's theory only refers to the psychological structure of people, but by extending this theory, it made it possible to use this theory for the study of various social, biological, economic, and... networks.
Inspired by the graph-theoretical framework, SBT has been developed as an approach to analyzing the organizational properties of complex networks \cite{antal2006social}. While graph theory was initially formulated to model dyadic relationships between information units, SBT extends beyond pairwise interactions to examine triadic relationships, offering deeper insights into network structures.\cite{moradimanesh2021altered}
When applying graph-theoretical analysis to brain networks, a graph G(V, E) is defined, where V represents a set of nodes corresponding to specific brain regions. The strength of association between each pair of regions of interest is then quantified as a set of weighted links (E), providing insight into the functional connectivity of the brain.
In SBT, the energy of a network ($E(N)$) measures its balance. The balance energy is calculated as shown in Equation~\ref{eq:balance_energy}. A positive product of edge signs within a triad indicates a balanced interaction, while a negative product signals imbalance. The energy is computed by summing the negative of these products across all triads in the network and normalizing by the total number of possible triads. Specifically, for a network with N nodes, the total number of possible triads is given by the combination formula $
\binom{N}{3}
$
. This normalization ensures that $E(N)$ is bounded within the range of $-1$ to $+1$, where $E(N) = -1$ corresponds to a fully balanced network and $E(N) = +1$ indicates a fully imbalanced one. The negative sign in the summation reflects the analogy to physical energy, where higher energy values are associated with greater frustration, and lower energy values correspond to more stable, balanced configurations.
\begin{equation}
E(N) = - \frac{1}{\binom{N}{3}} \sum_{i<j<k} s_{ij} \, s_{jk} \, s_{ik}.
\label{eq:balance_energy}
\end{equation}

\subsection*{Identification of Connected Imbalances Triads}

To investigate whether anti-synchronies are a global or local phenomenon, within the framework of SBT, we focus on frustrated triads in which all three links are negative. In other words, we focus on triple imbalances that reflect antisynchronous relationships between three regions of the brain that were obtained from the  400×400 brain network constructed in our work. This is because in imbalanced triads, where only one link is negative, it is not possible to observe the spread of antisynchrony across all three regions simultaneously. In such cases, the single negative link induces anti-synchrony only between two regions, while the third region remains synchronized with the others.
In contrast,  in imbalanced triads (with
three negative links), we can simultaneously observe the side effects of anti-synchrony: when anti-synchrony arises between two regions, it also appears in their interactions with the third region. Therefore, by examining the connectivity network of these fully negative triads, we can trace the spread of anti-synchrony and observe how it evolves with aging.
 We considered those triads whose balance energy exceeded the average balance energy of all $\Delta_{---}$ as $\Delta{^+}_{---}$. In addition, we restricted our analysis to cases in which such triads were connected, meaning that two  $\Delta_{---}$ shared a common edge. We denote these connected triads as $\Delta{^*}_{---}$.

\section*{Results and Discussion}
\subsection*{Whole-Brain Functional Correlation and Developmental Patterns}
\begin{figure}[H]
    \centering

    \subfloat[Probability Density Function of correlation across brain regions by age]{%
    \includegraphics[width=0.6\linewidth]{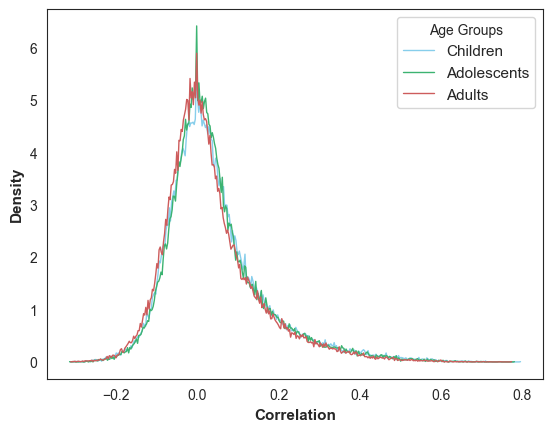}
    \label{fig:pdf}
 }
    \subfloat[ Adults]{%
        \includegraphics[width=0.4\textwidth]{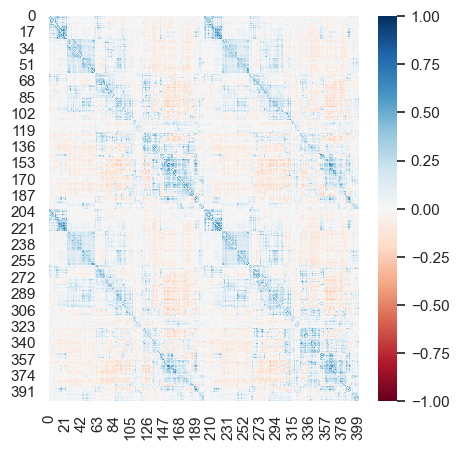}
        \label{fig:adultsheat}
    }
    
    \caption{(a) Probability Density Function of correlations among 400 brain regions across age groups. (b) Heatmaps of functional correlations for adults, showing subtle developmental refinements while maintaining overall stability in correlation structure.}
    \label{fig:heatmaps1}
\end{figure}

As highlighted in the introduction, the brain operates as a complex network, where not only individual regions but also the interconnections between them are essential for comprehending its functional and structural dynamics.
The Probability Density Function of correlations across 400 brain regions is illustrated by age groups in Fig.\ref{fig:pdf}. The KL divergence values—Children to Adolescents: 0.015, Children to adults: 0.013, and Adolescents to adults: 0.018—indicate minimal differences across age groups, suggesting that the changes in correlation distributions are not substantial or noticeable.
The heatmaps illustrate the correlation between brain regions. Due to the minimal differences between the KL divergence values across age groups and the absence of essential changes in the correlation distributions, we present only the heatmap for adults.
(see Fig.\ref{fig:adultsheat}) Symmetry in these correlations, particularly between regions in the left and the right hemispheres, indicates balanced inter-hemispheric communication.\\
Developmental differences are also evident in the heatmaps, with subtle variations in the intensity and distribution of correlations across age groups. These differences may reflect the maturation of functional networks or shifts in connectivity patterns as the brain develops. Despite these changes, the overall structure of brain connectivity remains relatively stable, as suggested by the minimal KL divergence values. This stability underscores that while developmental refinement occurs, the fundamental organization of the brain’s functional networks is preserved across age groups.
\\
For more detailed insights, we provide subnetwork-specific heatmaps that illustrate how specific regions within functional subnetworks, such as the Visual, Default mode, or Limbic networks, change over time. The reader can refer to the Appendix \ref{sec:appendixA} for additional information on these specific patterns and their implications.
\subsection*{The Distribution and Abundance of $\Delta_{---}$}
\begin{figure}[H]
    \centering

    \subfloat[Distribution Function of $\Delta_{---}$  by age groups]{%
    \includegraphics[width=0.6\linewidth]{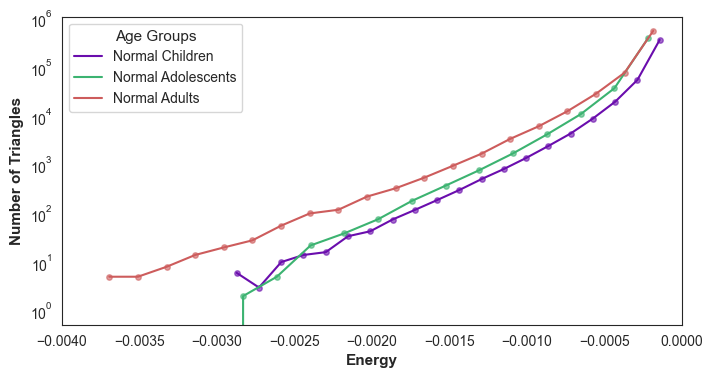}
    \label{fig:log}
 }
    
    \subfloat[ Children]{%
       \includegraphics[width=0.6\textwidth]{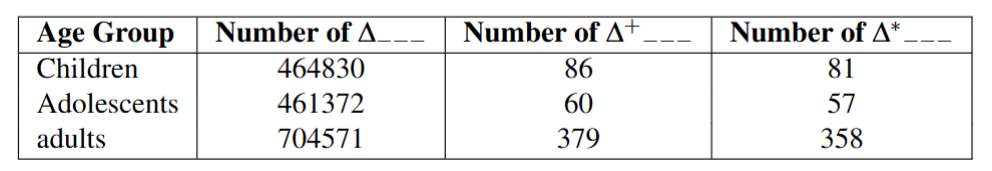}
       \label{fig:table}
    }
    \caption{a) Distribution function of $\Delta_{---}$  for children, adolescents, and adults. b) Exact counts of $\Delta_{---}$, $\Delta{^+}_{---}$, and $\Delta{^*}_{---}$ for each age group, highlighting developmental differences in balance energy profiles.}
    \label{fig:heatmaps1}
\end{figure}

In the previous section, we examined the correlation Probability Distribution Function (PDF) across different brain regions, as well as the functional connectivity map of 400 brain regions.
In this section we analyzed $\Delta_{---}$ the distribution function across different developmental stages. The distribution is plotted on a semi-logarithmic scale to capture variations across orders of magnitude more effectively Fig.\ref{fig:table}. The KL divergence values for these comparisons are as follows: children vs. adolescents: 0.041, children vs. adults: 0.001, and adolescents vs. adults: 0.020.
\\


The results indicate that while the overall distribution function remains consistent across age groups, there are variations in the intercepts. As illustrated in Fig.~\ref{fig:log}, the number of $\Delta_{---}$ increases as the balance energy approaches zero across all age groups. However, triads with lower balance energy are markedly fewer compared to those with small imbalance energy, which we previously denoted as $\Delta^{+}_{---}$. In our analysis, we focus on ($\Delta^{*}_{---}$) as we said before. Specifically, the mean balance energy of the Distribution Function of $\Delta_{---}$ was $-0.00151$ for children, $-0.00229$ for adolescents, and $-0.00194$ for adults; therefore, we selected connected $\Delta_{---}$ with balanced energy less than $-0.002$ as $\Delta^{*}_{---}$.
Figure~\ref{fig:table} presents the exact numbers of $\Delta_{---}$, $\Delta^{+}{---}$, and $\Delta^{+}{---}$ for each group.
\subsection*{Modularity and Network Organization of $\Delta^{*}_{---}$}
\begin{figure}[H]
    \centering
    \subfloat[ Children]{%
        \includegraphics[width=0.3\textwidth]{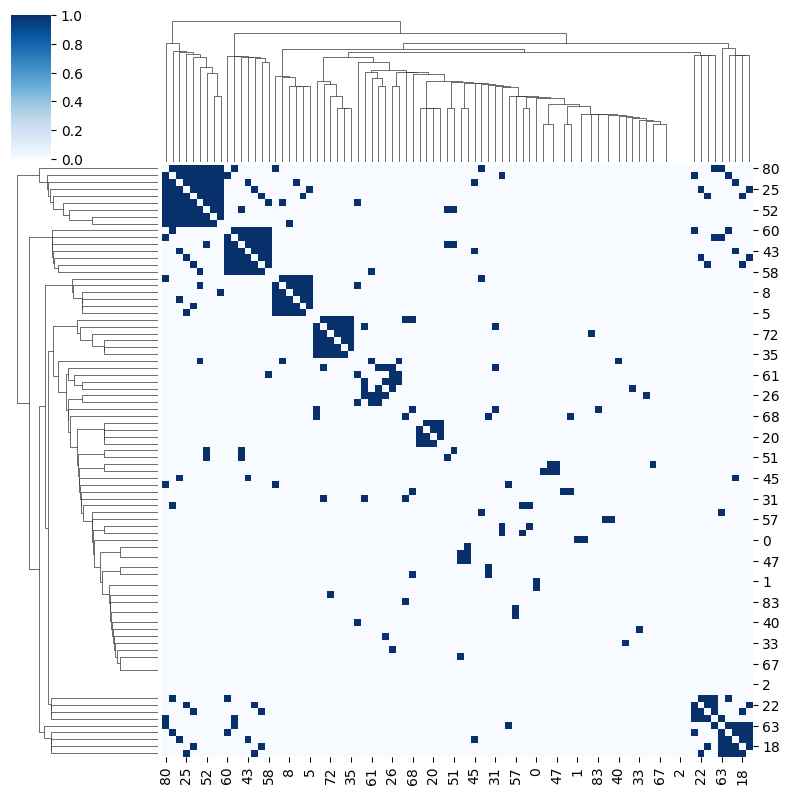}
        \label{fig:childrenden}
    }
    \subfloat[ Adolescents]{%
        \includegraphics[width=0.3\textwidth]{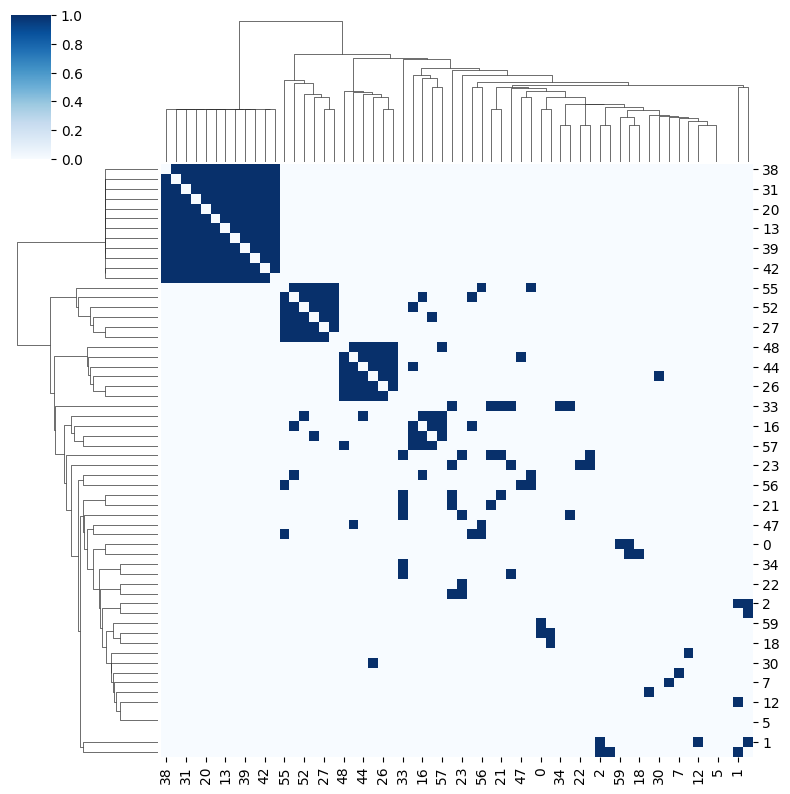}
        \label{fig:adolescentsden}
    }
    \subfloat[Adults]{%
        \includegraphics[width=0.3\textwidth]{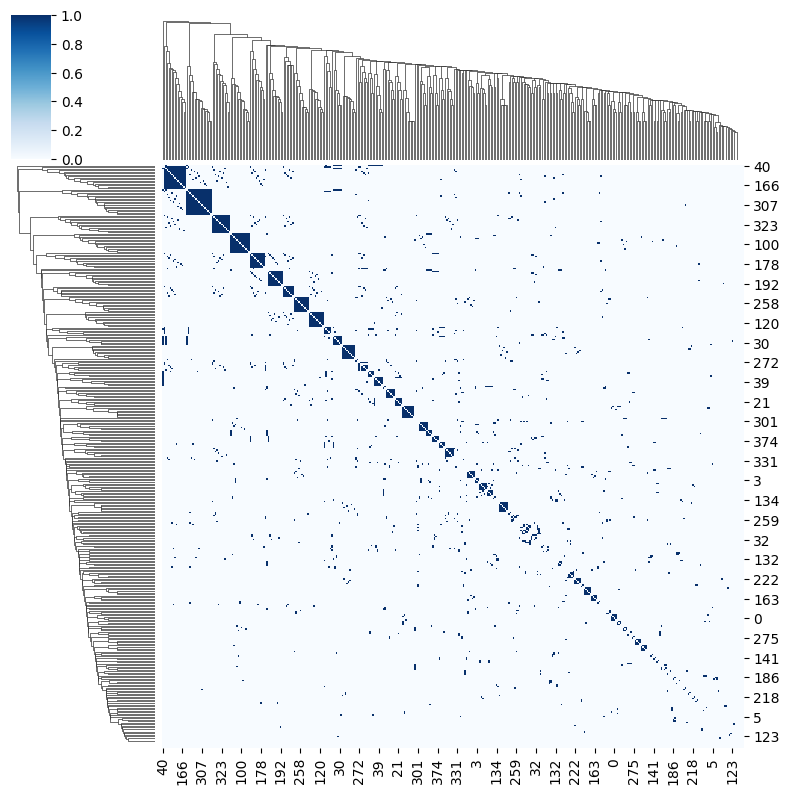}
        \label{fig:adultsden}
    }
    \hfill
    \subfloat[ Children]{%
        \includegraphics[width=0.38\textwidth]{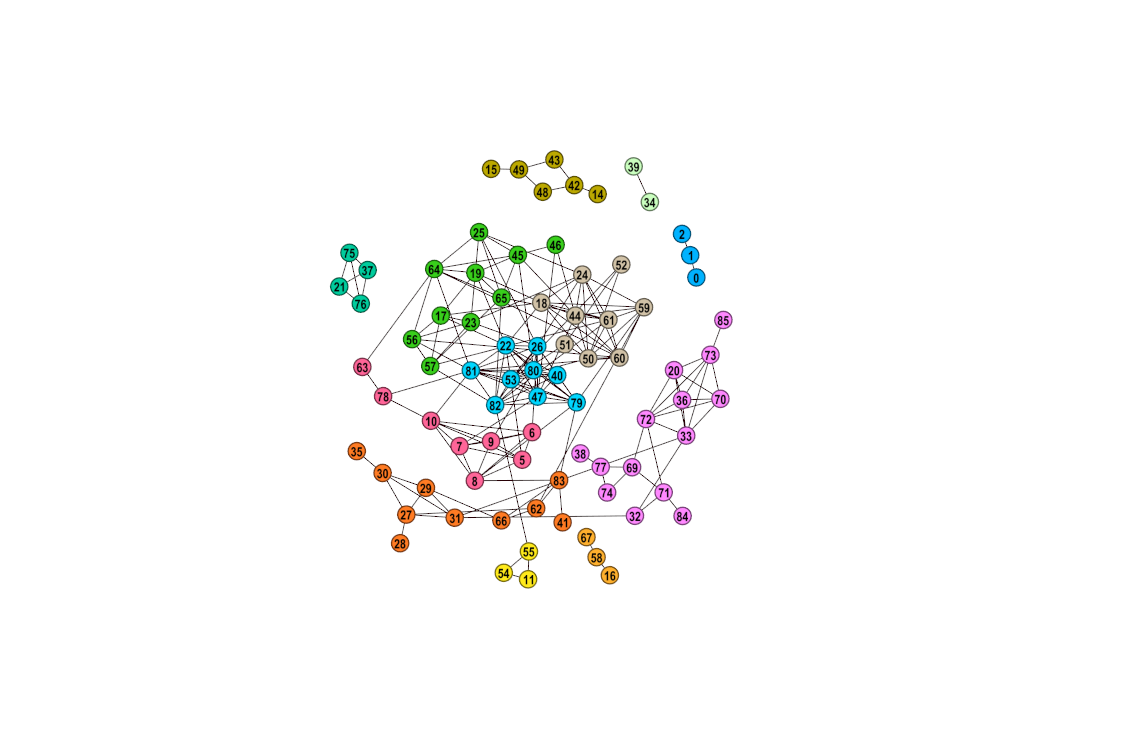}
        \label{fig:childrennet}
    }
    \subfloat[ Adolescents]{%
        \includegraphics[width=0.38\textwidth]{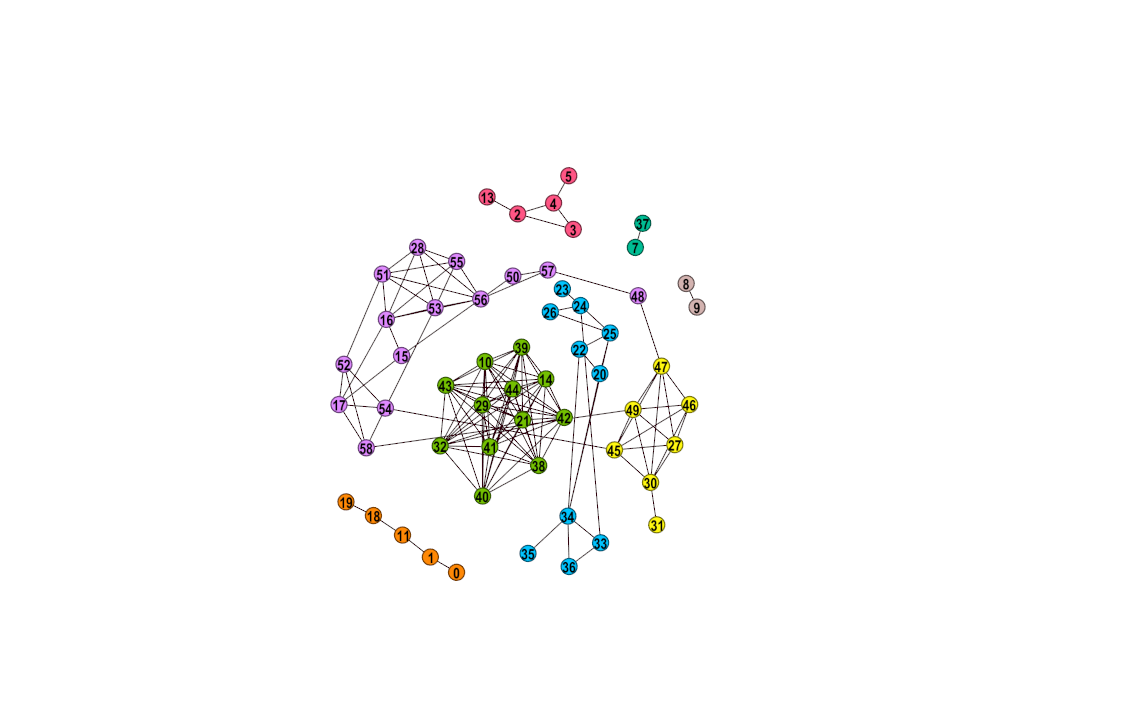}
        \label{fig:adolescentsnet}
    }
    \subfloat[Adults]{%
        \includegraphics[width=0.37\textwidth]{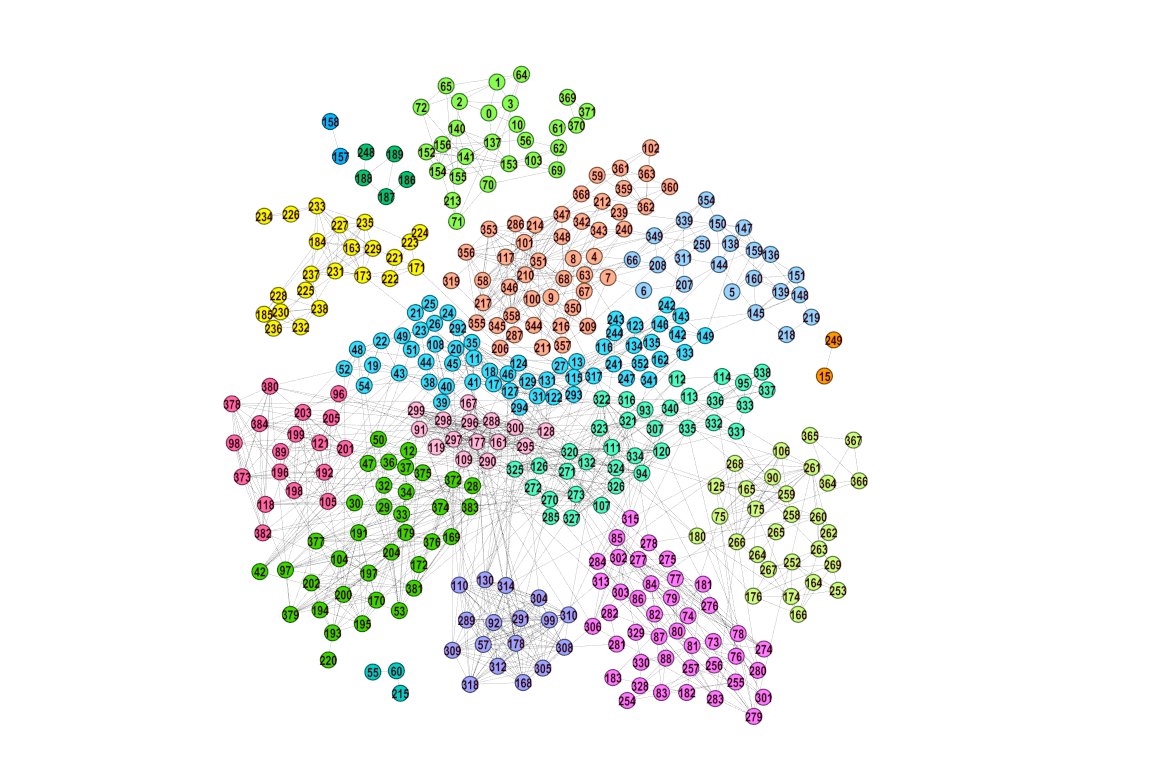}
        \label{fig:adultsnet}
    }
    
    \caption{(a–c) Dendrograms of $\Delta{^*}_{---}$ in children, adolescents, and adults, respectively. (d–f) Corresponding network visualizations reveal modularity and integration patterns, showing that $\Delta{^*}_{---}$ shifts from localized clusters in childhood to more interconnected structures in adulthood.}
    \label{fig:mainnetwork}
\end{figure}

To answer whether $\Delta{^+}_{---}$ is a local or global phenomenon, we first examine its dendrograms. The modular structures of their network are visible in the dendrograms (see Fig.\ref{fig:childrenden},\ref{fig:adolescentsden},\ref{fig:adultsden}). 
The dendrogram analysis reveals distinct developmental patterns in the organization of $\Delta{^*}_{---}$ across age groups. In adults, the network is significantly larger, with a high density of $\Delta{^+}_{---}$ distributed throughout, suggesting a more integrated but widespread presence of imbalances. The hierarchical clustering appears more continuous, with fewer clearly defined modules, indicating a more interconnected structure. In contrast, adolescence exhibits the fewest $\Delta{^+}_{---}$, with the largest group of imbalanced triads more distinctly separated from the others, suggesting a transitional phase in which the brain is reorganizing its network. This may reflect developmental processes such as synaptic pruning or shifts in functional connectivity, reducing local imbalances before integrating them into a more efficient and interconnected adult structure. Childhood presents a higher number of triads than adolescence, with strong modularity still visible, implying that imbalances are more confined within specific subsystems. The differences in sparsity and modular structure across age groups suggest that while $\Delta{^+}_{---}$ are initially localized in childhood, they become more distributed and integrated into the broader network by adulthood, reflecting a shift in the brain’s functional organization over development.
\\
To analyze the connection network between $\Delta{^+}_{---}$,as we mentioned we consider only $\Delta{^+}_{---}$ that are connected. In this network, each node represents a $\Delta{^*}_{---}$, and a link between two nodes indicates a common edge between the corresponding $\Delta{^*}_{---}$. We assign each $\Delta{^*}_{---}$ a numerical identifier that corresponds to its node number in the network (see Fig.\ref{fig:childrenden},\ref{fig:adolescentsnet},\ref{fig:adultsnet}).
\\
To visualize the functional networks we employed Gephi application\cite{bastian2009gephi,gephi}, an open-source platform for network analysis and graph-based data visualization. The layout was optimized using the Force Atlas-3D algorithm to enhance interpretability.
\\
Anti-correlation, or anti-synchrony, plays a crucial role in the organization and functionality of brain networks. In a foundational study by Fox et al. (2005)\cite{fox2005human}, it was demonstrated that the brain is intrinsically organized into dynamic, anticorrelated functional networks, particularly between the Default Mode Network (DMN) and task-positive networks. These anticorrelated patterns suggest that anti-synchrony is fundamental to the brain’s ability to switch between different cognitive states, thereby supporting efficient information processing and behavioral flexibility. In a more recent study, the importance of negative links was further emphasized in the context of emotional processing. The authors showed that negative connections contribute significantly to the structural balance and stability of brain networks during emotional experiences, highlighting their role in maintaining functional balance \cite{soleymani2023impact}. Together, these findings underscore that negative or anticorrelated interactions are not mere noise but essential components of healthy brain dynamics.
To investigate whether brain function driven by anticorrelations is a local or global phenomenon, we analyzed the network of $\Delta{^*}_{---}$. The structure of these networks across different age groups reveals that the brain exhibits specific organizational patterns in how $\Delta{^*}_{---}$ interacts.
An important question that arises from this network structure is whether the brain needs to coordinate and confirm all regional relationships to establish a connection, or whether it allows regions to act independently within local contexts.
\\
In childhood and adolescence, the $\Delta{^*}_{---}$ networks appear more discrete, indicating that brain functions operate more independently of one another during these developmental stages. The low KL divergence values between the distribution functions of $\Delta_{---}$ further support this interpretation, suggesting that there is no significant tendency for $\Delta_{---}$ to form connections with one another in these age groups. This lack of integration implies that decisions or functional changes within the brain are made locally, without the need for coordination or "permission" from the broader network. 
\\
In adulthood, however, the pattern is markedly different. The number of giant components in the $\Delta{^*}_{---}$ networks is noticeably higher compared to those in childhood and adolescence, indicating that a greater number of $\Delta{^*}_{---}$ are interconnected. This suggests that, unlike in earlier stages, the brain's functional organization in adulthood increasingly relies on the integration of imbalanced structures. During this developmental period, the function of the brain appears to emerge from a larger, more cohesive network of interacting $\Delta{^*}_{---}$, reflecting a shift toward global coordination and functional integration. 
The emergence of anti-synchrony between two brain regions does not affect only those regions—it can spread across the brain network, prompting a global response from the brain as a complex system. This propagation of anti-correlation can influence various parts of the system.
Consider a societal analogy: an increase in the dollar exchange rate leads to a rise in pesticide prices. Eventually, the price of milk also increases—not because the dollar is directly linked to milk, but because the rising cost of pesticides affects livestock management. Farmers cannot afford sufficient pesticides, their feed production declines, and as a result, milk production drops, leading to higher milk prices.
Similarly, in adulthood, the emergence of anti-synchrony between two brain regions does not remain localized. Other brain regions are also affected as part of the broader network response to this local disruption.
\\
In adulthood, brain function driven by the interactions between $\Delta{^*}_{---}$ triads emerges from a larger and more integrated network. Cognitive processes during this stage require greater coordination, and decisions tend to be made at a global level. This reflects a form of harmony and synchronization between the $\Delta{^*}_{---}$ and overall brain activity. At times, the adult brain network may even tolerate a small frustration if it benefits the system as a whole. In other words, each decision is evaluated collectively, and a final outcome is reached only when it is confirmed across the broader network. In adulthood, the brain appears to adopt a specific communication pattern for managing anticorrelations. The interactions between $\Delta{^*}_{---}$ triads serve as a framework for evaluating and guiding decision-making processes.
\subsection*{Energy–Energy Connectivity of $\Delta{^*}_{---}$}
\begin{figure}[H]
    \centering
    
    \subfloat[Children]{%
        \includegraphics[width=0.3\textwidth]{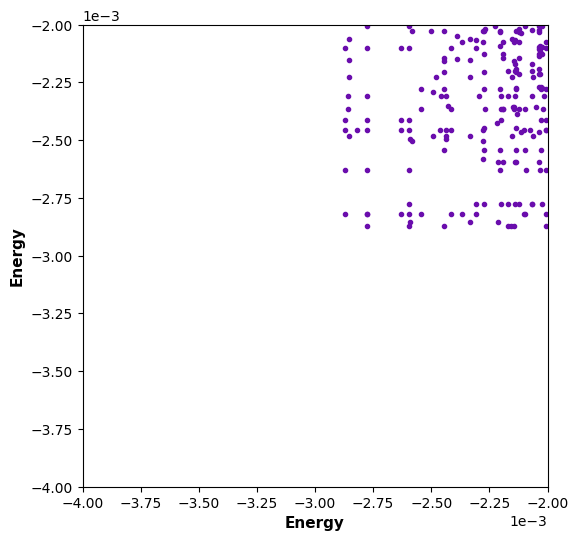}
        \label{fig:childrenee}
    }
    \hfill
    \subfloat[Adolescents]{%
        \includegraphics[width=0.3\textwidth]{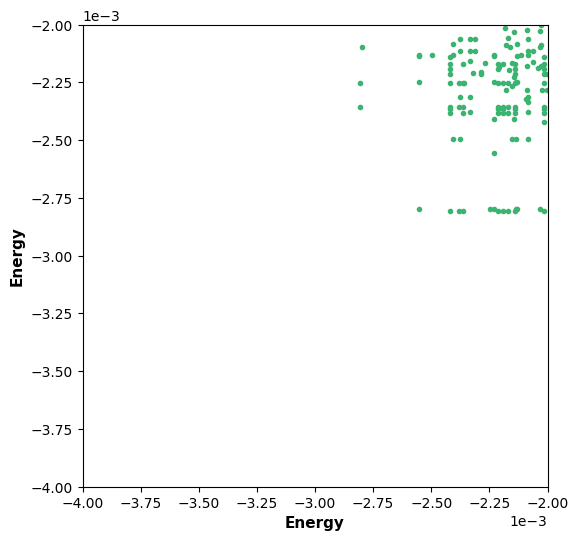}
        \label{fig:adolescentee}
    }
    \hfill
    \subfloat[Adults]{%
        \includegraphics[width=0.3\textwidth]{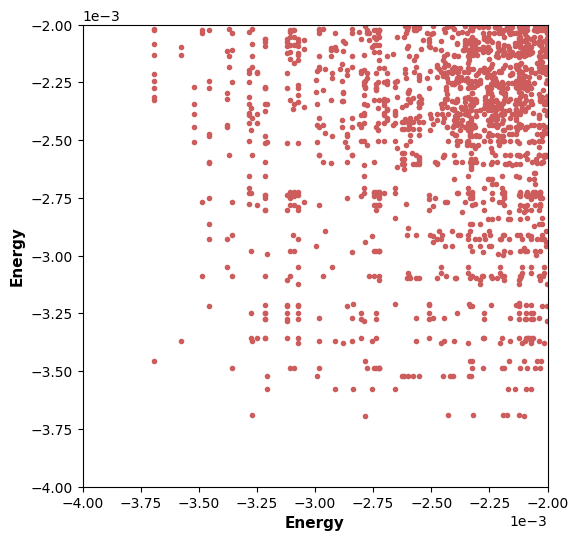}
        \label{fig:adultsee}
    }
    
    \caption{(a–c) energy–energy plots for children, adolescents, and adults. Results show that while only higher-energy imbalanced triads connect in childhood and adolescence, adulthood exhibits broad-spectrum connectivity, suggesting increased global integration of anticorrelations.}
    \label{fig:heatmaps}
\end{figure}

To identify which $\Delta{^*}_{---}$ are connected and the specific energy levels at which these connections occur, we analyze the energy-energy (e-e) graph.
The energy-energy (e-e) graphs highlight key developmental differences in the distribution and integration of $\Delta{^*}_{---}$ across age  (see Fig.\ref{fig:heatmaps}).
In childhood, the $\Delta{^*}_{---}$ that share common edges tend to have energies within a narrow interval, specifically between $-0.003$ and $-0.002$, indicating that only higher-energy structures are functionally interconnected at this stage. A similar pattern is observed during adolescence, despite a lower overall number of $\Delta{^*}_{---}$. In contrast, adulthood exhibits a markedly different organization: there is a strong synchronization between $\Delta{^*}_{---}$  across a broader energy spectrum. In this stage, both high- and low-energy $\Delta{^*}_{---}$ are functionally linked within the network, and their overall number is significantly increased. This suggests a more integrated and cohesive system in adulthood, where even imbalanced structures are incorporated into the broader network architecture.

\subsection*{Subnetwork-Level Anticorrelations Emerging from $\Delta{^*}_{---}$}
\begin{figure}[H]
    \centering
    
    \subfloat[Children]{%
        \includegraphics[width=0.25\textwidth]{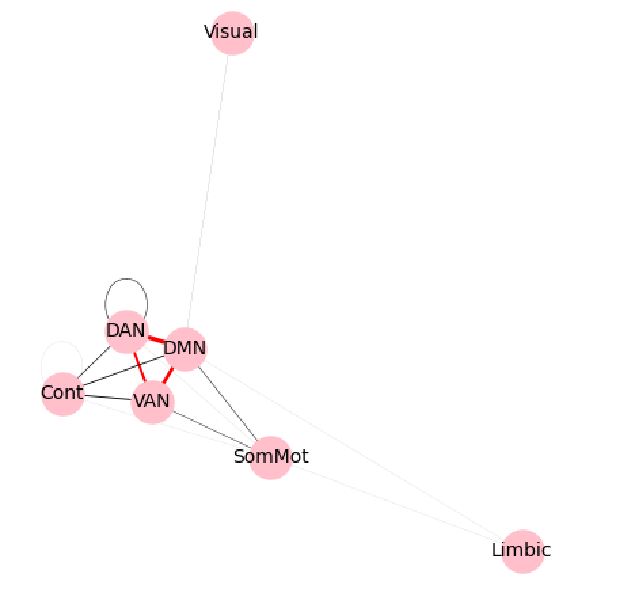}
        \label{fig:childrenee}
    }
    \hfill
    \subfloat[Adolescents]{%
        \includegraphics[width=0.25\textwidth]{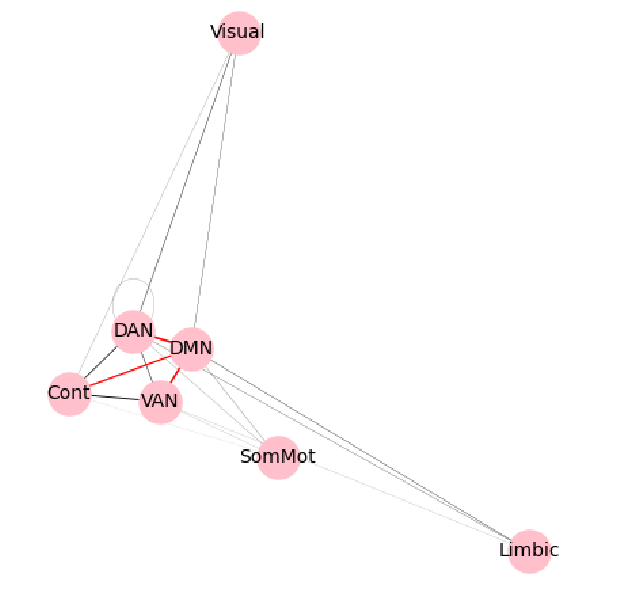}
        \label{fig:adolescentee}
    }
    \hfill
    \subfloat[Adults]{%
        \includegraphics[width=0.25\textwidth]{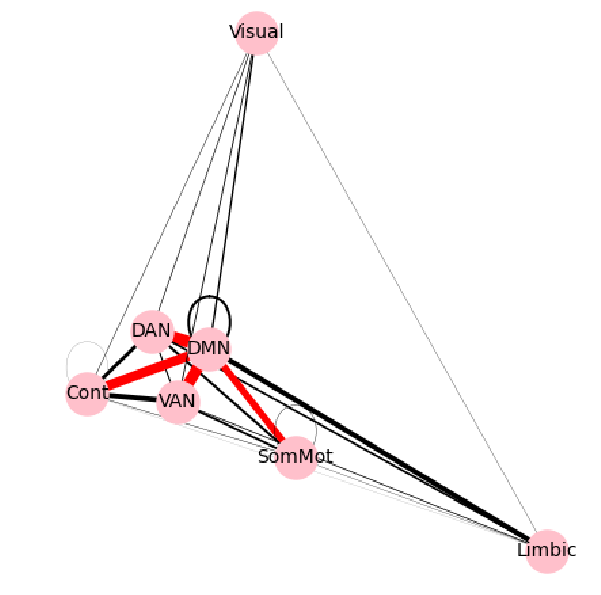}
        \label{fig:adultsee}
    }
    
    \caption{(a–c) Networks of anticorrelations between functional subnetworks in children, adolescents, and adults. Nodes represent subnetworks, while edge weights indicate the strength of negative connectivity. Results highlight the developmental trajectory from localized anticorrelations to strongly segregated and integrated subnetworks in adulthood.}
    \label{fig:subnet}
\end{figure}

We now aim to examine the $\Delta{^*}_{---}$  participating in Networks Fig.\ref{fig:childrennet},Fig.\ref{fig:adolescentsnet},Fig.\ref{fig:adultsnet} from a different perspective, taking a step back to observe the broader context. Each node in  these Networks represents a $\Delta{^*}_{---}$, with each of its vertices corresponding to one of the 400 distinct brain regions. We are particularly interested in identifying which brain subnetworks these vertices belong to—specifically, the subnetworks that are involved in the Networks.
To answer this question, we analyze all the nodes in the Networks. Since each node is a triad, each of its edges connects two brain subnetworks through an anti-synchronous (anticorrelated) relationship. To visualize the anti-synchronous interactions between the subnetworks, we define a new network, Fig.\ref{fig:subnet}. In this derived network, the nodes represent the brain subnetworks, and the links between them indicate the number of edges that connect the respective subnetworks negatively. Therefore, the higher the weight of a link, the more negative connections exist between the two subnetworks. 
Tab.\ref{t1}, Tab.\ref{t2}, and Tab.\ref{t3} present the connections with the highest weights between subnetworks, highlighting the links with considerable contributions to the overall network, for the children, adolescent, and adult groups, respectively.
It is important to emphasize that this derived network is extracted specifically from Networks Fig.\ref{fig:childrennet},Fig.\ref{fig:adolescentsnet},Fig.\ref{fig:adultsnet}. This means that it does not reflect all anticorrelations present in the brain but only those that form the edges of $\Delta{^*}_{---}$.
\\
A closer examination of anticorrelated subnetworks across development reveals a dynamic trajectory of functional segregation and integration that closely parallels cognitive maturation. Functional segregation refers to the extent to which regions within a given brain network are connected to each other, yet relatively connected to other networks. This supports specialized processing within domains such as attention, memory, and sensory input. In contrast, functional integration denotes the coordination and communication between distinct brain networks. This allows for flexible, goal-directed behavior and cross-domain cognitive control \cite{sporns2013network,wig2017segregated}. Optimal cognitive functioning requires balancing these two principles: segregation ensures efficiency and specialization, and integration enables adaptive behavior by flexibly combining information across systems.
In childhood, clear anticorrelations are already present, particularly between the Default Mode Network (DMN) and the Dorsal Attention Network (DAN), as well as between the DMN and the Salience/Ventral Attention Network (VAN). These early negative functional connections reflect the developing ability to differentiate between internally oriented thought (e.g., autobiographical memory, mind-wandering) and externally driven attentional systems \cite{fransson2007resting,gao2017functional}. The emergence of these patterns as early as age 2–4 suggests that functional segregation begins early, providing a neural foundation for attentional control and self-regulation in middle childhood \cite{fair2008maturing}.
In adolescence, however, this architecture undergoes a period of reorganization. While core anticorrelations—such as DMN–DAN—remain detectable, they are reduced in strength, consistent with a transient weakening of network boundaries due to synaptic pruning, white matter remodeling, and dynamic reconfiguration \cite{baum2017modular,marek2015contribution}. This “dip” in functional segregation during adolescence may support increased flexibility as the brain reorganizes and consolidates more efficient long-range connections. Importantly, these changes parallel behavioral shifts such as increased cognitive flexibility, emerging abstract reasoning, and adaptive emotional regulation \cite{luna2010has}.
By adults, the brain reaches a period of peak network segregation, where within-network connectivity is maximized and between-network anticorrelations are strongest \cite{chan2014decreased,nomi2017chronnectomic}. Notably, strong anticorrelations emerge not only between the DMN and DAN, but also between the DMN and the Frontoparietal Control Network (FPN), a configuration that supports efficient switching between internally generated thought and goal-directed behavior \cite{spreng2013intrinsic}. 
Other studies show that additional anticorrelations involving limbic, visual, somatomotor, and salience networks also become more prominent, suggesting greater modularity and cross-network coordination \cite{betzel2014changes}. 
These changes underpin adult-level cognitive functions such as sustained attention, working memory, and top-down executive control \cite{andrews2014default,satterthwaite2013heterogeneous}.
\\

\begin{table}[H]
\centering
\resizebox{0.6\textwidth}{!}{
\begin{tabular}{|l|l|c|c|}
\hline
\textbf{Node} & \textbf{Node} & \textbf{Weight} & \textbf{Weight / Total Weight} \\
\hline
\textbf{DorsAttn} & \textbf{Default} & 68.0 & 0.138 \\
\textbf{SalVentAttn} & \textbf{Default} & 55.0 & 0.112 \\
\textbf{DorsAttn} & \textbf{SalVentAttn} & 41.0 & 0.083 \\
\hline
\end{tabular}
}
\caption{Major anticorrelated subnetworks identified from $\Delta{^*}_{---}$ in childhood. The listed weights reflect the strength of anticorrelation between subnetworks, with proportions showing their relative dominance within the network.}
\label{t1}
\end{table}

\begin{table}[H]
\centering
\resizebox{0.6\textwidth}{!}{
\begin{tabular}{|l|l|c|c|}
\hline
\textbf{Node} & \textbf{Node} & \textbf{Weight} & \textbf{Weight / Total} \\
\hline
\textbf{DorsAttn}   & \textbf{Default}   & 34.0 & 0.162 \\
\textbf{SalVentAttn}   & \textbf{Default}   & 31.0 & 0.148 \\
\textbf{Cont}   & \textbf{Default}  & 26.0 & 0.124 \\

\hline
\end{tabular}
}
\caption{Major anticorrelated subnetworks identified from $\Delta{^*}_{---}$ in adolescence. The listed weights reflect the strength of anticorrelation between subnetworks, with proportions showing their relative dominance within the network.}
\label{t2}
\end{table}

\begin{table}[H]
\centering
\resizebox{0.6\textwidth}{!}{
\begin{tabular}{|l|l|c|c|}
\hline
\textbf{Node} & \textbf{Node} & \textbf{Weight} & \textbf{Weight / Total Weight} \\
\hline
\textbf{DorsAttn} & \textbf{Default} & 192.0 & 0.238 \\
\textbf{SalVentAttn} & \textbf{Default} & 178.0 & 0.221 \\
\textbf{Cont} & \textbf{Default} & 155.0 & 0.192 \\
\textbf{SomMot} & \textbf{Default} & 111.0 & 0.138 \\

\hline
\end{tabular}
}
\caption{Major anticorrelated subnetworks identified from $\Delta{^*}_{---}$ in adulthood. The listed weights reflect the strength of anticorrelation between subnetworks, with proportions showing their relative dominance within the network.}
\label{t3}
\end{table}

\clearpage
\appendix
\section{}
\begin{figure}[!htb]
    \centering
    \subfloat[Children]{%
        \includegraphics[width=0.3\textwidth]{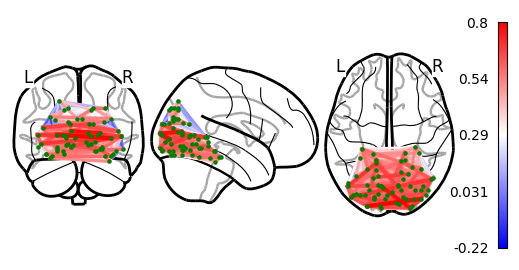}
    }
    \hfill
    \subfloat[Adolescents]{%
        \includegraphics[width=0.3\textwidth]{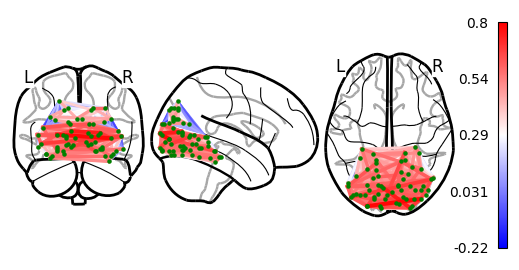}
    }
    \hfill
    \subfloat[adults]{%
        \includegraphics[width=0.3\textwidth]{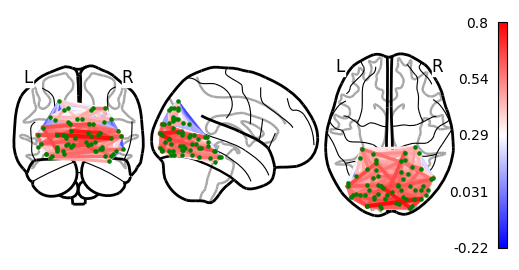}
    }
    \caption{Correlations across the Visual subnetwork regions for different age groups: a) children, b) adolescents, c) adults.}
    \label{fig:visual}
\end{figure}

\begin{figure}[!htb]
    \centering
    \subfloat[Children]{%
        \includegraphics[width=0.3\textwidth]{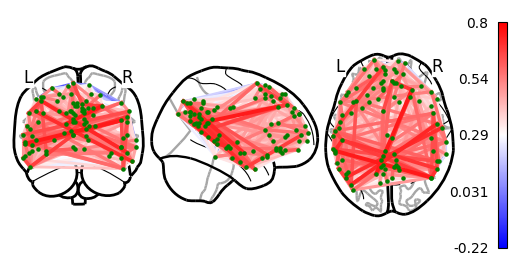}
    }
    \hfill
    \subfloat[Adolescents]{%
        \includegraphics[width=0.3\textwidth]{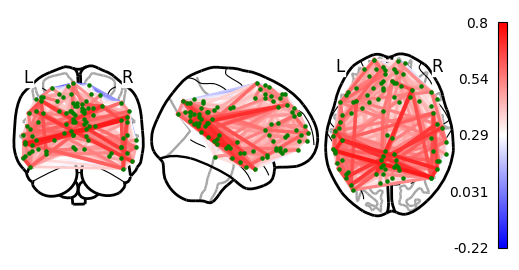}
    }
    \hfill
    \subfloat[adults]{%
        \includegraphics[width=0.3\textwidth]{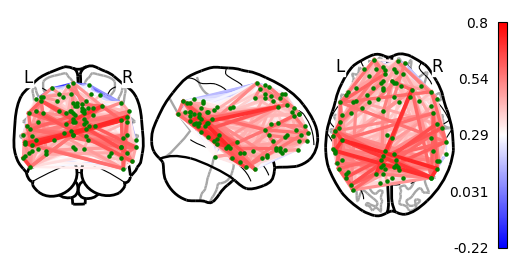}
    }
    \caption{Correlations across the Default mode subnetwork regions for different age groups: a) children, b) adolescents, c) adults.}
    \label{fig:Deafault}
\end{figure}

\begin{figure}[!htb]
    \centering
    \subfloat[Children]{%
        \includegraphics[width=0.3\textwidth]{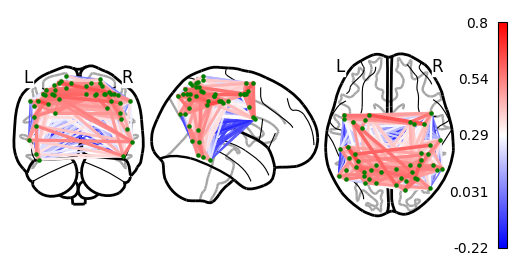}
    }
    \hfill
    \subfloat[Adolescents]{%
        \includegraphics[width=0.3\textwidth]{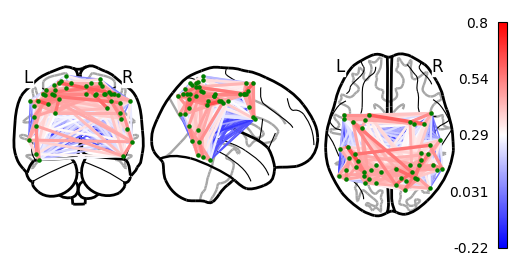}
    }
    \hfill
    \subfloat[adults]{%
        \includegraphics[width=0.3\textwidth]{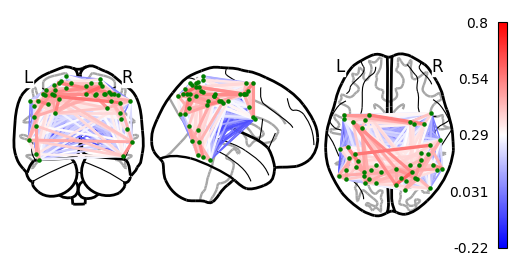}
    }
    \caption{Correlations across the DorsAttention subnetwork regions for different age groups: a) children, b) adolescents, c) adults.}
    \label{fig:DorsAttn}
\end{figure}

\begin{figure}[!htb]
    \centering
    \subfloat[Children]{%
        \includegraphics[width=0.3\textwidth]{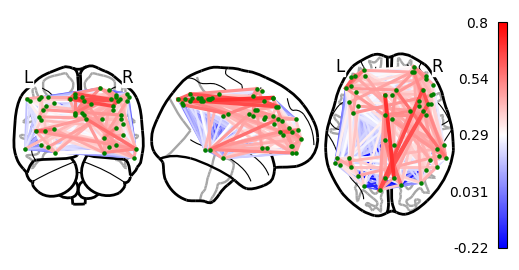}
    }
    \hfill
    \subfloat[Adolescents]{%
        \includegraphics[width=0.3\textwidth]{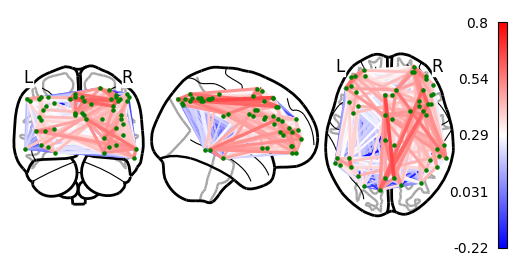}
    }
    \hfill
    \subfloat[adults]{%
        \includegraphics[width=0.3\textwidth]{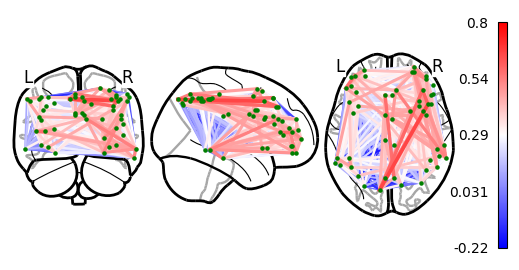}
    }
    \caption{Correlations across the Cont subnetwork regions for different age groups: a) children, b) adolescents, c) adults.}
    \label{fig:Cen}
\end{figure}

\begin{figure}[!htb]
    \centering
    \subfloat[Children]{%
        \includegraphics[width=0.3\textwidth]{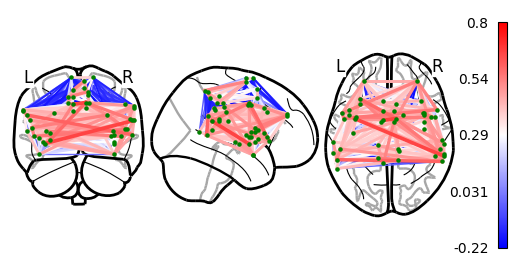}
    }
    \hfill
    \subfloat[Adolescents]{%
        \includegraphics[width=0.3\textwidth]{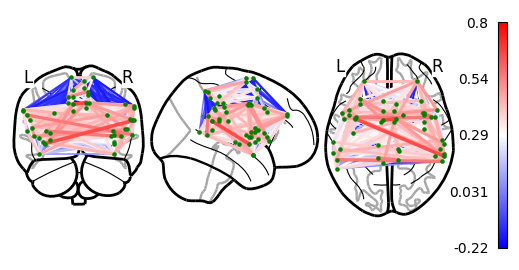}
    }
    \hfill
    \subfloat[adults]{%
        \includegraphics[width=0.3\textwidth]{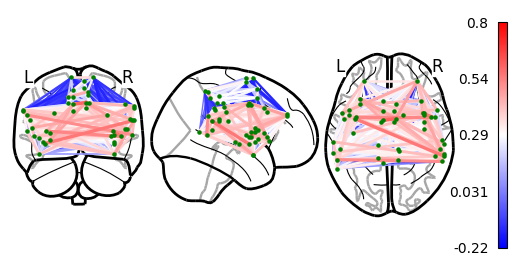}
    }
    \caption{Correlations across the SalvAttention subnetwork regions for different age groups: a) children, b) adolescents, c) adults.}
    \label{fig:salvAttn}
\end{figure}

\begin{figure}[!htb]
    \centering
    \subfloat[Children]{%
        \includegraphics[width=0.3\textwidth]{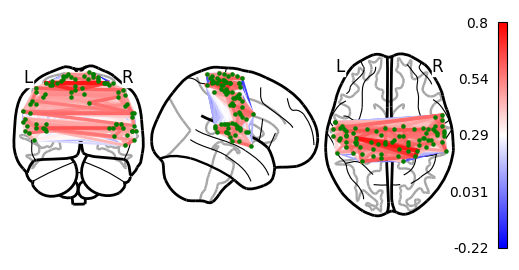}
    }
    \hfill
    \subfloat[Adolescents]{%
        \includegraphics[width=0.3\textwidth]{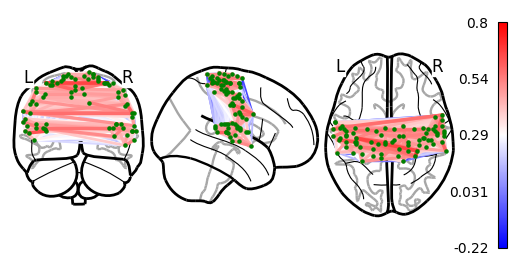}
    }
    \hfill
    \subfloat[adults]{%
        \includegraphics[width=0.3\textwidth]{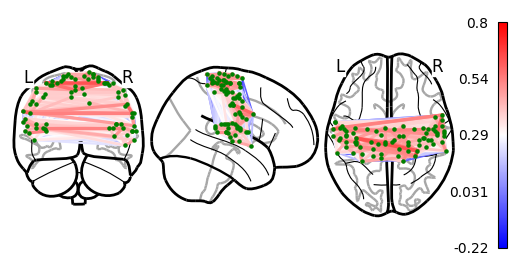}
    }
    \caption{Correlations across the Somoto Motor subnetwork regions for different age groups: a) children, b) adolescents, c) adults.}
    \label{fig:samMort}
\end{figure}

\begin{figure}[!htb]
    \centering
    \subfloat[Children]{%
        \includegraphics[width=0.3\textwidth]{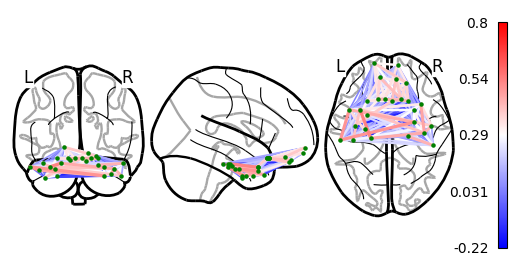}
    }
    \hfill
    \subfloat[Adolescents]{%
        \includegraphics[width=0.3\textwidth]{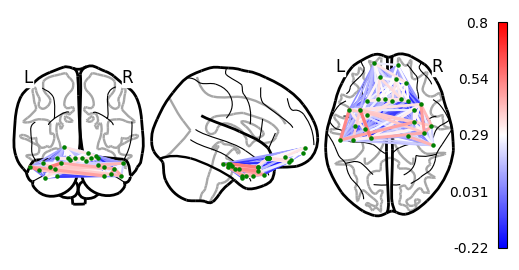}
    }
    \hfill
    \subfloat[adults]{%
        \includegraphics[width=0.3\textwidth]{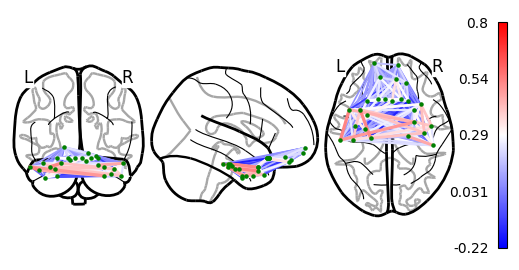}
    }
    \caption{Correlations across the Limbic subnetwork regions for different age groups: a) children, b) adolescents, c) adults.}
    \label{fig:Limbic}
\end{figure}

For consistency and clear comparison, the color scales for the correlation heatmaps were standardized across all subnetworks, with the maximum correlation value set at 0.8 and the minimum at -0.22, corresponding to the highest and lowest observed pairwise correlations in the whole-brain connectivity matrix. While maximum and minimum correlations within each subnetwork may vary, this uniform scale allows for meaningful visual comparison of connectivity patterns across different networks and developmental stages.
These graphs provide a detailed visualization of the within-subnetwork correlation patterns across childhood, adolescence, and adulthood. The heatmaps Fig.\ref{fig:visual},Fig.\ref{fig:Deafault},Fig.\ref{fig:DorsAttn},Fig.\ref{fig:Cen},Fig.\ref{fig:salvAttn},Fig.\ref{fig:samMort},Fig.\ref{fig:Limbic} illustrate how functional connectivity among regions within each specific subnetwork changes over development.
For the Visual Network Fig.\ref{fig:visual}, correlations remain strong and stable across all age groups, suggesting consistent integration of visual processing regions throughout development.
The Default Mode Network (DMN) Fig.\ref{fig:Deafault} shows a high degree of within-network correlation in childhood, with a subtle decrease in adolescence, possibly reflecting synaptic pruning and refinement processes. In adulthood, correlations within the DMN appear more stable, suggesting a mature, functionally cohesive network supporting self-referential and introspective processes.
In the Dorsal Attention Network (DAN) Fig.\ref{fig:DorsAttn}, within-network correlations are moderate in childhood and adolescence but appear more diffuse in adulthood, possibly reflecting the DAN's dynamic role in task engagement and reorientation of attention.
The Control Network (CEN) Fig.\ref{fig:Cen} exhibits relatively sparse within-network correlations in all age groups, with slight increases in adulthood, potentially reflecting the ongoing development of executive functions.
The Salience Network Fig.\ref{fig:salvAttn} demonstrates modest within-network coherence across development, suggesting a stable but less tightly integrated structure compared to other networks.
For the Somatomotor Network Fig.\ref{fig:samMort}, strong within-network correlations are evident in childhood and persist into adulthood, reflecting motor functions' stable, fundamental role.
Lastly, the Limbic Network Fig.\ref{fig:Limbic} shows weaker and more variable within-network correlations, especially in adolescence and adulthood, which may reflect the complex and flexible nature of emotional processing systems.

\label{sec:appendixA}

\clearpage
\bibliography{references}
\bibliographystyle{abbrv}

\section*{Author contributions statement}
Conceptualization: Zahra Moradimanesh, Reza Khosrowabadi, G.Reza Jafari
Data Creation:  Zahra Moradimanesh, Fahimeh Ahmadi. 
Formal analysis: Fahimeh Ahmadi.
Investigation: Fahimeh Ahmadi.
Methodology: Fahimeh Ahmadi, G.Reza Jafari.
Project administration: G.Reza Jafari.
Visualization: Fahimeh Ahmadi
Writing – original draft: Fahimeh Ahmadi, G.Reza Jafari.
Writing – review \& editing: Fahimeh Ahmadi, Zahra Moradimanesh, and Reza Khosrowabadi, G.Reza Jafari.

\end{document}